# Expertise-Based Developer Assignment for Long-Term Software Components in Open-Source Projects

Faheem Ullah[1], Babar Shah[2], William Shanks[3], Ayesha Mohsin[4], Abrar Ullah[5]

[1]College of Interdisciplinary Studies, Zayed University, Abu Dhabi, UAE
[2]College of Technological Innovation, Zayed University, Abu Dhabi, UAE
[3]University of Adelaide, Adelaide, South Australia, Australia
[4]National University of Computer and Emerging Sciences, Islamabad,
[5]School of Mathematical and Computer Sciences, Heriot-Watt University, Dubai

**Abstract.** Open-source software development through GitHub has enabled countless software projects to be developed by developers from across the world. Assigning the right task to the right developer enables teams to work efficiently. Assigning a task which is not representative of a developer's expertise results in slower completion time and less maintainable code. Whilst much work has been done to automate this assignment for specific jobs, there is little work addressing the assignment of developers to expertise-specific long-term components of a new project. This paper produces a web application that models developers' expertise based on their previous Git commits and automatically assigns them to an optimal task within a new project. Testing showed that the system's speed varies depending on the back-end model used but server-side data caching improved worst-case speeds by a factor of 9.86. 72.4% of tasks had their target developer listed in the top 10, out of a possible 47.

## 1. Introduction

GitHub is the largest collaborative version-control platform for open-source software development, hosting contributions from over 40 million developers across nearly 200 million projects (Khillar, 2021). Git allows crucial insight into the expertise of any developer as their activity can be tracked across projects (Waldenberg et al., 2017). This has allowed projects to draw on a global pool of contributors. The success of open-source development hinges on matching the right developer to the right task. Open-source projects that allow any developer to self-assign tasks can result in them working in areas misaligned with their skillsets (Cooper, 2020). This leads to contributions that fail to meet project standards and are consequently rejected. Mature projects built around cohesive and familiar teams address this by having project maintainers manually assign tasks based on their impression of each contributor's strengths. Whilst this improves productivity as compared to self-assignment, maintainers are rarely exposed to a contributor's complete body of work. So, these assign-ments are inevitably based on partial information and are vulnerable to bias (Terrell et al., 2017).
Prior work has attempted to automate parts of this assignment process. Bug triaging systems have routed individual bugs to developers using developer vocabulary and self-reported skills (Yadav et al., 2019). These approaches work on well-defined and short-lived tasks with an existing internal project history. Less attention has been paid to assigning developers to long-term, expertise-specific components at the start of a project. Prior works have either relied on consultation-based methods (André et al., 2011), or were built for paid task marketplaces instead of GitHub (Mao et al., 2015). There is no existing system that automatically assigns developers to long-term project roles on GitHub according to their expertise. The paper addresses this gap by developing a system to allocate developers to long-term, expertise-specific components using only the developer's cross-project commit history. The objective is to ensure that tasks demanding a technical skillset are matched with the developers best suited to fulfil them. A web application is built in which a project maintainer enters a list of candidate developers and a list of tasks including their required libraries. The system then draws on a vector representation of developer and library expertise to compute a percentage match between every developer and every task.
The paper is organised as follows. Section 2 reviews prior literature. Section 3 proposes the general methodology to produce the system. Section 4 explores the experimental setup of the system. Section 5 then shows the results of the two parameters tested. Section 6 discusses all findings from the report. Section 7 concludes the paper and proposes future work.

## 2. Related Work

### 2.1. Text-mining Open-source Repositories

(Le et al., 2021) utilised topic modelling on over 70,000 security vulnerability posts on Stack Overflow and Security Stack Exchange to define 13 distinct topics discussed within the field. The approach was extremely effective at modelling human language. However, software code has a completely different structure and syntax, so modelling expertise would require alternate approaches.
In a large-scale review of big data, (Gandomi and Haider, 2015) stated that while topic modelling is the most utilized in text analysis, more unconventional approaches may suit analysis of software development. These include adjusting the text analysis methodologies to suit the unique characteristics of code, coding syntax, and file metadata. (Ma et al., 2021) developed a framework for such methodologies called World of Code (WoC). The framework is generated from over 70 million GitHub projects and presents data in a way where the authors, projects, commits, blobs and dependencies can be cross-tracked and filtered. It is updated routinely to reflect the entire GitHub. The size of the project makes it infeasible to reproduce and given its availability for academic reuse, this WoC framework is considered a key resource in our project.

### 2.2. Modelling Developer Expertise

(Ullah et al., 2026) modelled cybersecurity professionals' skill profiles by mining 12,161 job ads and 49,002 Stack Overflow posts showing that textual data can reliably proxy expertise. (Dey et al., 2021) built a subset of the WoC data where each developer, project, library and programming language is represented as a vector in vector space. This enables calculations between two entities to measure their similar usage, such as cosine similarity (Han et al., 2012). (Santos et al., 2021) used libraries and application programming interfaces (APIs) within code files to inform a model that automatically labels bugs found in development. They assumed that a developer could decide whether they can fix an issue if they can see the skillset required. (Montandon et al., 2019) combined GitHub commit activity with self-reported skills on LinkedIn and trained supervised machine learning (ML) classifiers to predict expertise. (Wan et al., 2018) proposed SCSMiner, which mines social coding platforms simultaneously and propagates relevance scores across a constructed graph. Although using evidence from multiple platforms improved precision over single-source models, we restrict our system to GitHub commit data only. Multiple platforms would require developers to maintain external profiles which our system cannot assume are available.

### 2.3. Auto-Assigning Developers to Tasks

Much of the current research of assigning developers revolves around bug triaging. (Matter et al., 2009) modelled the vocabulary of developers' source code and assigned bugs based on a match between developers' vocabulary and the bug reports' vocabulary.

(Yadav et al., 2019) optimised systems for automatically allocating bugs to developers by developer expertise scores (DES) based on priority and average fix-time, which provided a priority ranking for each developer to each bug. This feature is highly prioritised in our work as it provides the project manager with a readout of their top options to make an informed decision (Xiao et al., 2023). (Aung et al., 2022) extended this by proposing a multi-task learning framework that assigns a developer and classifies issue type from a single model. They used a text encoder alongside an abstract syntax tree encoder to jointly represent bug descriptions and code.

For developer allocation across an entire project, (André et al., 2011) developed a Delphi-based expert consultation method for assigning developers to projects. However, they utilized a psychological framework instead of an analytical software approach. (Mao et al., 2015) addressed role assignment on TopCoder where developers are paid to complete tasks from random projects rather than on GitHub.

Given its broader utilisation, GitHub is a valuable platform to develop our intended model around. The idea behind bug triaging projects is that bugs are already somewhat classified. Further ML, meta-data analysis, and processing are a means to improve these auto-assignment models. However, trying to auto-assign developers to entire tasks of the software project requires more foresight. Our approach of building a model around commits across all projects they've been involved with on GitHub increases the training data size and therefore assignment accuracy. Bringing the concepts of modelling a developer's expertise based on the libraries/APIs they have previously used will inform the role auto-assignment system developed in this project.

# 3. Proposed Methodology

The system begins as a web application with a Node.js instance handling all background processing before returning results to the front end. The system outputs percentage matches between each developer and each task, based on analysis of the developers' prior contributions. After system implementation, the system is subjected to a structured evaluation phase. This includes accuracy, which is measured against a real-world historical case study; and efficiency, which is examined through systematic variation of the number of developers, the number of tasks, and the number of libraries specified per task. The complete system code is presented in (Shanks et al., 2022).

(Dey et al., 2021) Skill Space is utilised extensively in this work. The vector space serves as the basis for the library-extraction mechanism of (Shanks et al., 2021) that this project builds on. If candidate developers are represented in the Skill Space, similarity calculations can be made directly between them and the task libraries giving the most accurate possible match. If input developers are not represented in the Skill Space, then the system pulls their commits from WoC and runs the text analysis for modelling their expertise from scratch. This is because Skill Space is trained on only a subset of the WoC framework. The extracted library names are used to retrieve corresponding library representations from the Skill Space for similarity calculations.

## 3.1. System Implementation

The back end was implemented in Python because of its suitability for modular, segmented development. Initial development was conducted in a Jupyter Notebook with scripts subsequently exported to native Python files for execution on the Node.js server. Pandas was utilised for data handling. Work was carried out on a machine with a 2.2 GHz Quad-Core Intel Core i7 processor and 16 GB of 1600 MHz DDR3 RAM.

Seen from **Fig. 1**, the system involves three core parts: the front-end web interface, a base server and the WoC server. The front end is delivered as a standard HTML, CSS and JavaScript package from the base server upon HTTP request. It offers fields for a project name, a set of tasks each with a title and a line-separated list of libraries, and a list of available developers. Clicking "Submit" triggers "submit()", which then utilises the built-in JavaScript DOM functionality to scrape the fields from the page, sanitise the input for unwanted whitespace, and send the data as stringified JSON in the header of a POST request to the server. The base server, running on Node.js, manages the incoming request and coordinates the back-end processes required to produce the output.

From here, developers are split into two groups: those already represented in the Skill Space, and those who are not. For the first group, compatibility with a task is comput-ed as the cosine similarity between the developer's vector and the vector representation of each library in that task's set, with the mean similarity across a task's libraries taken as the developer's percentage match. For the second group, the system connects to the WoC server via RSA-encrypted SSH and runs a remotely stored Python script with the developer's name as a parameter, gathering code from up to 10,000 of that developer's Git commits (Mombach and Valente, 2018) and sending it back to the base server via SSH FTP. This code then passes through the authors' library-extraction framework, which parses it for library names in use; the Skill Space representations of those libraries are aggregated to stand in for the developer's own vector. From there, similarity calculations proceed exactly as they do for Skill Space-represented developers, again averaged across libraries.

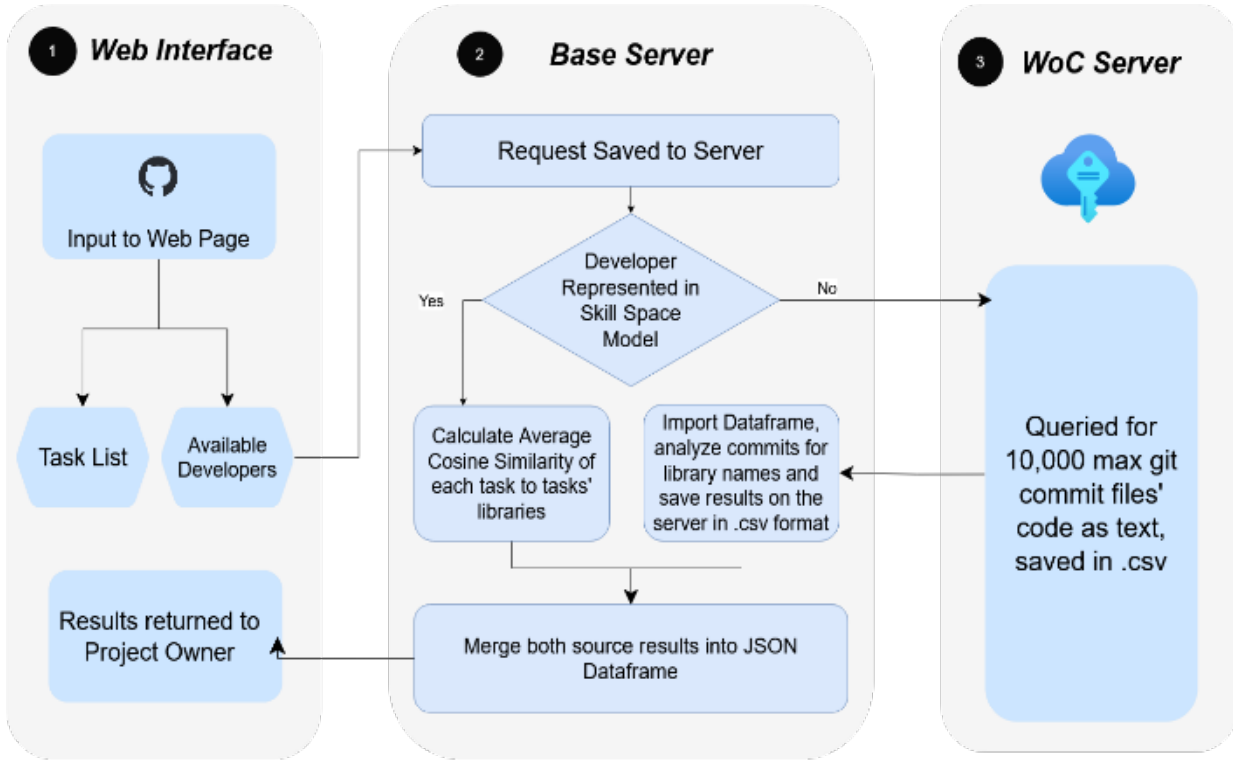

**Fig. 1.** System Diagram

### 3.2. Accuracy Analysis

The Eclipse bug dataset is a collection of bug instances gathered across a decade of development on Eclipse open-source products. This dataset has been commonly utilised to validate whether bug-triaging solutions assign issues to the correct developer. This dataset was not a good fit for our purposes for two reasons. First, it does not use Git for version control, and second, it specifies bugs instead of the long-term tasks our system is designed around. However, many Eclipse projects have since migrated to GitHub and this migration offered a workable alternative. By analysing the developers who contributed most actively to Eclipse's most active project, deeplearning4j, the full set of libraries in use across that project was extracted. If each of those libraries is fed into the system as individual tasks, and all the developers are input as candidate options, then it can be observed which developer the system judged the best match for each library. By tallying up which developer utilised each library the most, the system's predictions against this measure of "most prevalent user" as a narrow, library-level stand-in for expertise can be compared.

### 3.3. Scalability Analysis

(Ullah et al., 2022) evaluated big data system scalability by varying parameters systematically and analysing performance as distributions rather than single readings. Using this as a basis, the scalability assessment of our system takes a controlled approach built around three variables: the number of candidate developers (devSize), the number of tasks defined (taskSize), and the number of libraries associated with each task (librarySize).

To assess how the system speed responds to changes in these variables, a sample dataset consisting of 19 developers, 9 tasks, and 5 libraries per task was defined. A Python script utilising BeautifulSoup and Selenium libraries opened new instances of the system in a controlled browser. Input data was chosen at random to initiate the process. On the server side, the same script that manages the back-end calculation records the duration of each run with a resolution of $0.1 \times 10^{-6}$ seconds of accuracy. Each run picks a different combination of developers, tasks, and libraries per task to get a large range of readings at different permutations. To account for the risk of results being distorted by arbitrary configuration, the resulting measurements are treated as distributions rather than single readings. Time recording starts only after the Skill Space model has been loaded into memory because loading it takes a consistent, unavoidable 12.45 seconds on every system run.

Modelling a developer's expertise from scratch, connecting to the WoC server, waiting for the remote script to execute, transferring results back via FTP, and performing text analysis is a slow process. So, the system caches the final output of every WoC developer on the base server so that any subsequent query of that same developer can skip the process entirely. The scalability test was then run across three separate contexts: with developers already in the Skill Space (the fastest), with developers not in the Skill Space but that have been cached from previous queries (slower), and the non-Skill Space developers without caching (slowest). All benchmarking was carried out with the Node.js server and the browser automation script running on the same machine.

## 4. Experimental Results

### 4.1. Accuracy Evaluation

Accuracy testing evaluated whether the similarity mechanism could predict which developer working on deeplearning4j was indeed the greatest match for that expertise. Here, a lower rank indicates greater predictive accuracy. Of the 50 libraries, one was assigned with perfect accuracy, with the target developer placed first. A further 10 libraries placed the target developer within the top 5 of 47 developers, and 26 more placed them within the top 10. The remaining 14 libraries placed the target developer within the top 50%, with three outliers landing at positions 24, 26, and 42. **Fig. 2** shows that system rankings of the target developer skewed toward the top of the list across most libraries. 72.4% of the libraries placed the target developer within the top 10, which is a meaningful indicator of system accuracy. These results are a reasonable indication that the system can differentiate developers' expertise at the library level and correctly assign tasks to these skilled individuals who may not have otherwise been assigned the task from longer traditional manual analysis.

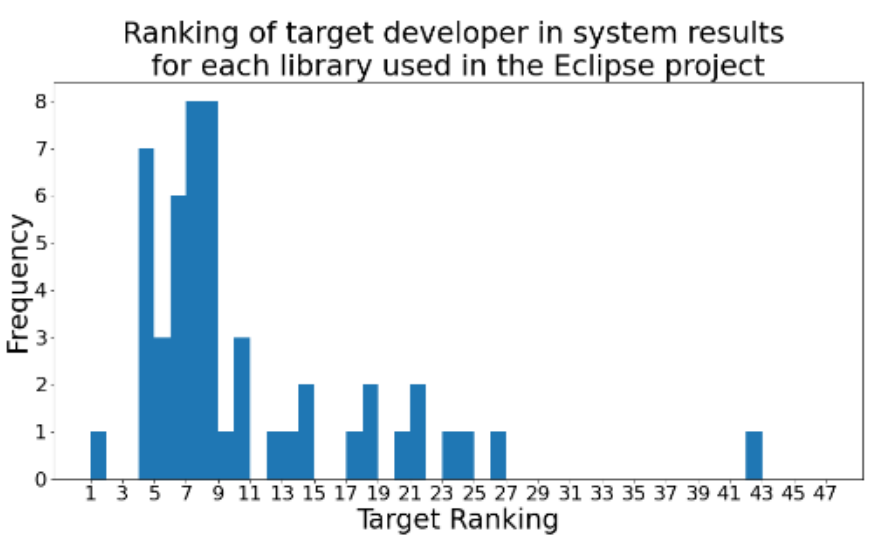

**Fig. 2.** Eclipse target rank within system ranking

### 4.2. Scalability Evaluation

Scalability was first assessed under the condition that all candidate developers had representations within Skill Space. The results are presented as boxplot distributions with each box depicting the minimum, 25th percentile, mean, 75th percentile, and maximum modelling times at each discrete parameter value. As seen in **Fig. 3**, increasing any of the three parameters increases modelling time on a linear regression since these developers can be accessed directly from memory. The dominant computational cost lies in the similarity calculations. Each incremental increase in the variables increases the number of times these calculations are performed, reflected in an increased modelling time. Absolute modelling times remained extremely low, from 0.07 to 0.16 seconds even in the worst case, coming from lightweight looping arithmetic.

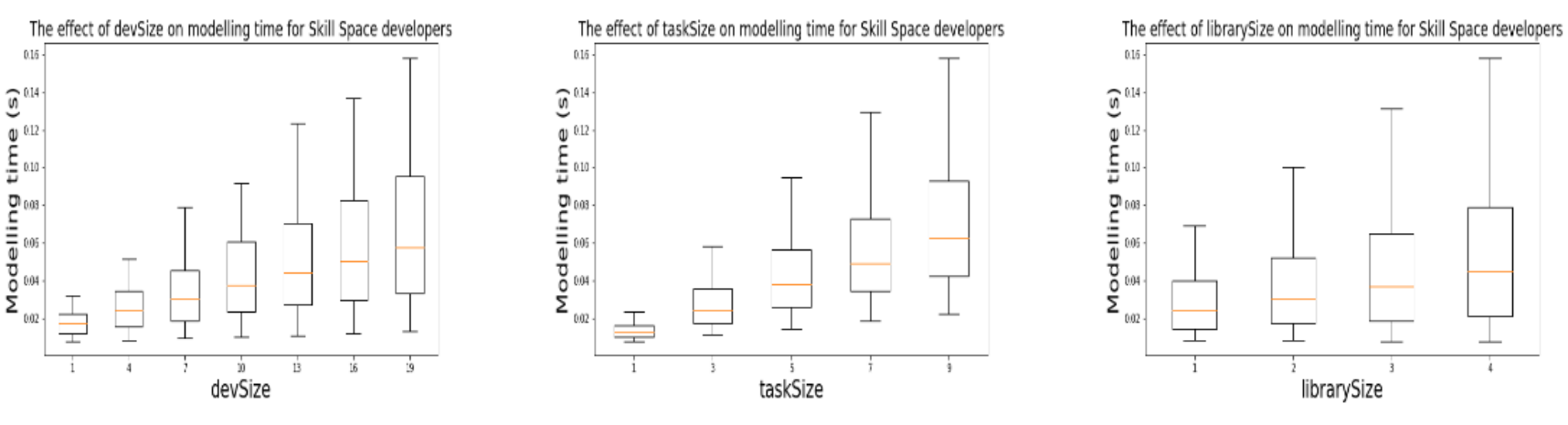

**Fig. 3.** The effect on modelling time for Skill Space developers

**Fig. 4** shows the modelling time in the case of non-cached WoC developers. A large increase in modelling time is due to networking with the WoC server and performing text analysis, pushing the worst-case time to between 100 and 300 seconds. However, taskSize and librarySize have a negligible effect. This is because these variables only add further loop iterations to the calculations, which are fast once the developers and tasks are already in vectors. The distributions here rarely overlap at each discrete value. This suggests that the time taken to retrieve each developer from the WoC servers is uniform and consistent.

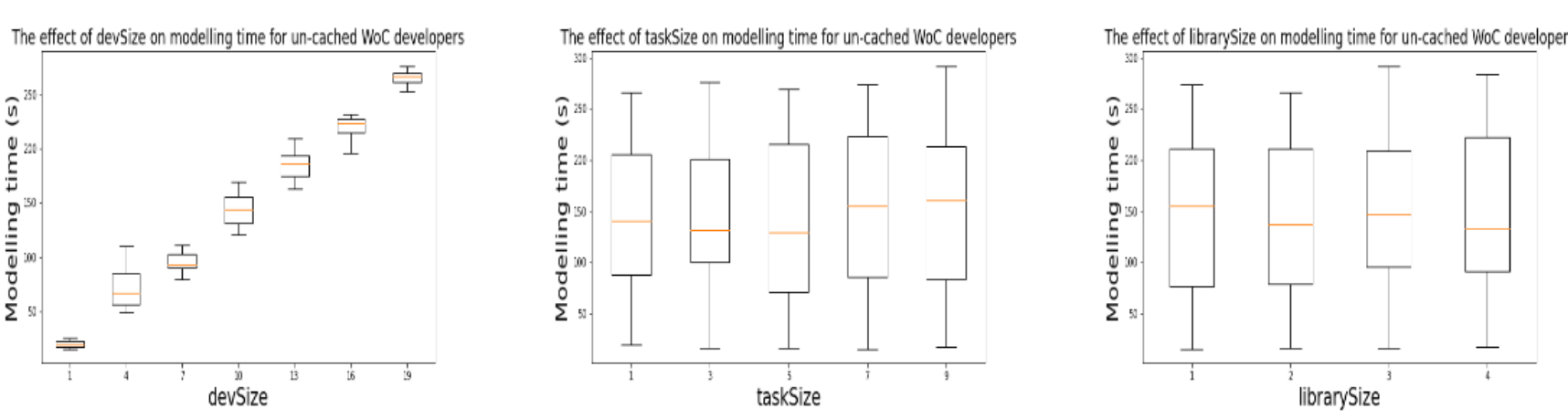

**Fig. 4.** The effect on modelling time for non-cached WoC developers

**Fig. 5** shows results for developers outside the Skill Space but whose results have been cached to the server. Modelling times are non-trivial but large enough that all three variables influence the outcome. Caching reduces worst-case modelling time by a factor of up to 9.86 relative to an uncached query.

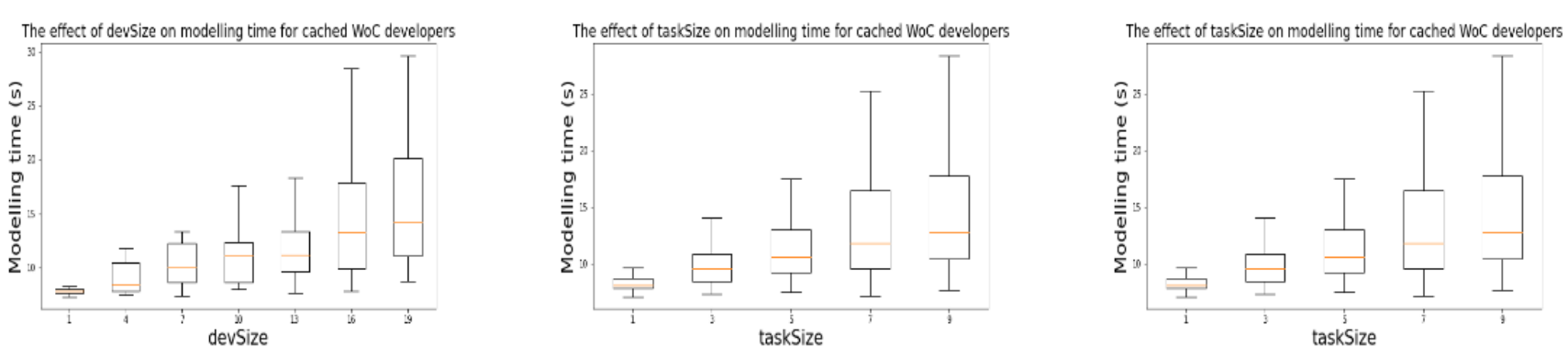

**Fig. 5.** The effect on modelling time for cached WoC developers

## 5. Discussion

This paper is not the first to automate developer-role assignment, but it is the first to combine full automation, a purely commit-derived model, and GitHub deployment within a single system. (André et al., 2011) assigned roles through human judgement rather than computation. (Mao et al., 2015) took a computational approach within TopCoder. Our system differs by combining full automation with a model built entirely from developers' own commit history on the platform where open-source collaboration most commonly takes place. The results support (Dey et al., 2021) Skill Space that developers and libraries can be positioned within a shared vector space.

These results reveal three distinct patterns. First, clusters near the top of the ranking rather than across the developer pool imply that the model is capturing real structure in commit history. Second, since project maintainers draw on an overlapping pool of contributors across related projects rather than a fresh set each time, most real-world queries would benefit from caching after an initial run. This would push performance toward the faster end rather than the worst case. Third, that devSize dominates scalability across every condition indicates the bottleneck lies in comparing across developers rather than in task complexity. This is a promising sign for scaling to larger projects. Controlling a multi-layered JavaScript system also introduces some timing variability from asynchronous calls and network operations that these results cannot fully rule out.
The findings have practical implications for open-source project managers. By automating task-to-developer allocation based on the developer's commit history, project maintainers can assign more accurately. This mitigates the bias that comes with partial information, and they can consider developers encountered elsewhere but not known personally. Integrating such a system with platforms like GitHub is a reasonable direction for reducing friction in the task-assignment-heavy early phase of a new project. It also has the potential downstream benefit of fewer bugs and lower maintenance burden as tasks are better matched to expertise.

### 5.1. Limitations

First, accuracy evaluation is testing a proxy for expertise rather than expertise itself. Since Eclipse's bug-tracking data was not directly usable, we substituted the most prolific user of a library as a stand-in for the correct assignee. This may conflate usage frequency with task fit. Second, scalability testing is limited by a modest testing sample size. Third, all testing was carried out with the Node.js server and the browser automation script running on the same local machine. This was essential for device security and minimising connection variability. Therefore, reported timings may reflect local, near-zero-latency conditions rather than the network and server latency the system would face once deployed. Finally, whilst the most common open-source system is Git, Subversion and Mercurial could be integrated into a future version (Kemper, 2012).

## 6. Conclusion

This project assigns developers within an open-source project to the optimal task, based on their expertise, according to code previously contributed. A web application is developed allowing project owners of new open-source projects to input their tasks, the libraries relevant, and the developers available, and receive a quantitative measure of how well each developer fits each task. Testing against a real-world example, the system recommends developers to library-specific tasks with 72.4% of tasks having their target developer listed in the top 10 out of a possible 47 options. When testing scalability, developers in the Skill Space result in near-instant modelling, whilst those requiring WoC result in far greater times. Importantly, caching speeds this up by a factor of 9.86. The number of developers is the most influential factor for scalability. For future work, role-level assignment accuracy should be tested using genuine long-term task outcomes. On the modelling side, incorporating artificial intelligence (AI) techniques to interpret a project owner's plain-language description of a task would lower the barrier to entry for non-technical maintainers who may not always know which specific libraries or skills a task involves. Similarly, keeping the Skill Space resident in server memory between requests would offer a straightforward route to improved execution speed.

## 7. References

1. Khillar, S.: Difference Between GitHub and SourceForge. http://www.differencebetween.net/technology/difference-between-github-and-sourceforge/, last accessed 2022/03/16 (2021)
2. Terrell, J., et al.: Gender differences and bias in open source: pull request acceptance of women versus men. PeerJ Comput. Sci. 3, e111 (2017)
3. Cooper, Z.: Getting started with contributing to open source. https://stackoverflow.blog/2020/08/03/getting-started-with-contributing-to-open-source/, last accessed 2022/03/16 (2020)
4. Waldenberg, A., van Andel, B., Kastner, C.: Gitinspector. https://github.com/ejwa/gitinspector, last accessed 2022/03/27 (2017)
5. Le, T.H.M., et al.: A large-scale study of security vulnerability support on developer Q&A websites. In: Evaluation and Assessment in Software Engineering, pp. 109–118. Trondheim, Norway (2021)
6. Gandomi, A., Haider, M.: Beyond the hype: big data concepts, methods, and analytics. Int. J. Inf. Manag. 35(2), 137–144 (2015)
7. Ma, Y., et al.: World of code: enabling a research workflow for mining and analyzing the universe of open source VCS data. Empir. Softw. Eng. 26(2), 22 (2021)
8. Dey, T., Karnauch, A., Mockus, A.: Representation of developer expertise in open source software. In: 2021 IEEE/ACM 43rd International Conference on Software Engineering (ICSE), pp. 995–1007 (2021)
9. Santos, F., et al.: Can I solve it? Identifying APIs required to complete OSS tasks. In: 2021 IEEE/ACM 18th International Conference on Mining Software Repositories (MSR) (2021)
10. Ullah, F., Babar, M.A.: On the scalability of big data cyber security analytics systems. J. Netw. Comput. Appl. 198, 103294 (2022)
11. Matter, D., Kuhn, A., Nierstrasz, O.: Assigning bug reports using a vocabulary-based expertise model of developers. In: 2009 6th IEEE International Working Conference on Mining Software Repositories (MSR), pp. 131–140 (2009)
12. Yadav, A., Singh, S.K., Suri, J.S.: Ranking of software developers based on expertise score for bug triaging. Inf. Softw. Technol. 112, 1–17 (2019)
13. André, M., Baldoquín, M.G., Acuña, S.T.: Formal model for assigning human resources to teams in software projects. Inf. Softw. Technol. 53(3), 259–275 (2011)
14. Mao, K., et al.: Developer recommendation for crowdsourced software development tasks. In: 2015 IEEE Symposium on Service-Oriented System Engineering (2015)

15. Shanks, W., Ullah, F.: Topics in Comp Sci - First Release, v1.0.0. https://github.com/FaheemCrest/William_2022/releases#release-v0.1.0 (2021)
16. Mombach, T., Valente, M.T.: GitHub REST API vs GHTorrent vs GitHub Archive: a comparative study (2018)
17. Han, J., Kamber, M., Pei, J.: Getting to Know Your Data, pp. 39–82. Elsevier, Amsterdam (2012)
18. Kemper, C., Oxley, I.: Foundation Version Control for Web Developers. Apress, New York (2012)
19. Shanks, W., Ullah, F.: Automatic Assignment of Git Users to Tasks in New Project v0.1.0. https://github.com/FaheemCrest/William_2022, last accessed 2022/05/30 (2022)
20. Xiao, W., Li, J., He, H., Qiu, R., Zhou, M.: Personalized first issue recommender for newcomers in open source projects. In: 2023 38th IEEE/ACM International Conference on Automated Software Engineering (ASE), pp. 800–812 (2023)
21. Montandon, J.E., Lourdes Silva, L., Valente, M.T.: Identifying experts in software libraries and frameworks among GitHub users. In: 2019 IEEE/ACM 16th International Conference on Mining Software Repositories (MSR), pp. 276–287 (2019)
22. Aung, T.W.W., Wan, Y., Huo, H., Sui, Y.: Multi-triage: a multi-task learning framework for bug triage. J. Syst. Softw. 184, 111133 (2022)
23. Wan, Y., Chen, L., Xu, G., Zhao, Z., Tang, J., Wu, J.: SCSMiner: mining social coding sites for software developer recommendation with relevance propagation. World Wide Web 21(6), 1523–1543 (2018)
24. Ullah, F., Ye, X., Fatima, U., Wu, Y., Akhtar, Z., Ahmad, H.: What skills do cybersecurity professionals need? Inf. Comput. Secur. 1–19 (2026)